\documentclass[%
 reprint,
 superscriptaddress,
 preprintnumbers,
 amsmath,amssymb,
 aps,
floatfix,
]{revtex4-2}

\usepackage[utf8]{inputenc}
\usepackage[T1]{fontenc} 
\usepackage{graphicx}
\usepackage{dcolumn}
\usepackage{bm}
\usepackage{xcolor}

\renewcommand{\arraystretch}{1.2}
\makeatletter
\newcommand{\mathleft}{\@fleqntrue\@mathmargin0pt}
\newcommand{\mathcenter}{\@fleqnfalse}
\makeatother

\begin{document}

\preprint{ADP-19-24/T1104}
\preprint{DESY 19-190}
\preprint{Liverpool LTH 1217}

\title{Electromagnetic contribution to $\Sigma$--$\Lambda$ mixing using lattice QCD+QED}

\author{Z.R.~Kordov}
\affiliation{%
 CSSM, Department of Physics, University of Adelaide, SA, Australia
}%
\author{R.~Horsley}
\affiliation{School of Physics and Astronomy, University of Edinburgh, Edinburgh EH9 3FD, UK}
\author{Y.~Nakamura}
\affiliation{RIKEN Center for Computational Science, Kobe, Hyogo 650-0047, Japan}
\author{H.~Perlt}
\affiliation{Institut für Theoretische Physik, Universit\"at Leipzig, 04109 Leipzig, Germany}
\author{P.E.L.~Rakow}
\affiliation{Theoretical Physics Division, Department of Mathematical Sciences,University of Liverpool, Liverpool L69 3BX, UK}
\author{G.~Schierholz}
\affiliation{Deutsches Elektronen-Synchrotron DESY, 22603 Hamburg, Germany}
\author{H.~St\"uben}
\affiliation{Regionales Rechenzentrum, Universit\"at Hamburg, 20146 Hamburg, Germany}
\author{R.D.~Young}%
\affiliation{%
 CSSM, Department of Physics, University of Adelaide, SA, Australia
}%
\author{J.M.~Zanotti}
\affiliation{%
 CSSM, Department of Physics, University of Adelaide, SA, Australia
}%

\collaboration{CSSM/QCDSF/UKQCD Collaboration}
\noaffiliation

\date{\today}

\begin{abstract}
Mixing in the $\Sigma^0$--$\Lambda^0$ system is a direct consequence of broken isospin symmetry and is a measure of both isospin-symmetry breaking as well as general SU(3)-flavour symmetry breaking. In this work we present a new scheme for calculating the extent of $\Sigma^0$--$\Lambda^0$ mixing using simulations in lattice QCD+QED and perform several extrapolations that compare well with various past determinations. Our scheme allows us to easily contrast the QCD-only mixing case with the full QCD+QED mixing. 
\end{abstract}

\maketitle


\section{\label{sec:introduction}Introduction\protect\\}

Our best theoretical understanding of the classification and qualities of the low-lying hadron states comes from the theory of SU(3)-flavour symmetry, first summarized by the eight-fold way \cite{GellMann:1961ky}. Since SU(3)-flavour is an approximate symmetry, broken by non-degeneracy in the physical properties of the up, down and strange quarks, it is often only a convenient starting point for precision determinations of hadron properties.

The neutral $\Sigma$ and $\Lambda$ states of the spin-$1/2$ baryon octet, as defined by SU(3)-flavour, differ only in isospin (also denoted T-spin \cite{deSwart:1963pdg}) which is not an exact symmetry in nature. Consequentially the physical particle states that correspond to these octet baryons are actually mixtures of the idealised isospin-states. 
They are the only two states in the baryon octet that have the same quark content and charge and thus permit mixing under isospin-breaking. The neutral members of the pseudoscalar meson nonet present an analogous system of mixing (see e.g. \cite{Dudek:2011tt,Ottnad:2019xmh}. 

The amount of mixing that occurs in the physical system depends on the degree of isospin-symmetry breaking, as well as further SU(3)-flavour symmetry breaking by the strange quark, and is driven by both the bare mass parameters and the differences in the quark charges.

One approach in calculating the $\Sigma^0$--$\Lambda^0$ mixing, due to Dalitz and Von Hippel (denoted DvH \cite{Dalitz:1964es}), is based on relationships between the electromagnetic mass-splittings of octet baryons. This was derived by consideration of an effective Lagrangian density exhibiting SU(3)-flavour symmetry plus a perturbation which encodes bare quark mass and QED effects \cite{Coleman:1961jn}, and uses experimental baryon masses as inputs. 

Our approach herein is to use simulations in lattice QCD+QED to fit a parametrization of the $\Sigma^0$--$\Lambda^0$ mixing angle, which we derive by considering the effects of continuous variations in the quark mass and charge parameters around an SU(3)-symmetric point, and extrapolate to physical values of the quark masses and charges.
We find that the inclusion of QED effects in our determination gives us a result comparable to that of DvH, which is the only other determination to explicitly include electromagnetic effects.

Further, when QED is ignored we show that our new extrapolation scheme gives good agreement with previous QCD-only calculations of the mixing, using chiral perturbation theory ($\chi$PT) \cite{Isgur:1979ed,GASSER198277} and lattice QCD (no QED \cite{Horsley:2014koa,Horsley:2015lha}), and gives a magnitude of about half of that found by DvH or this work when QED is incorporated. 

In Section~\ref{sec:SLmixing} we introduce the practical structure for
probing $\Sigma^0$--$\Lambda^0$ using lattice QCD+QED and 
derive a parametrization for use in extrapolating our 
lattice results to the physical point (physical quark 
masses and QED-coupling). 
We give details of the lattice simulation parameters used in Section~\ref{sec:lattice} before performing the extrapolation and presenting our results in Section~\ref{sec:results}, and we also observe some traits of the 
mixing in the cases of QCD-only and QCD+QED and contrast these cases. We conclude this work in Section~\ref{sec:conclusion}.

\section{\label{sec:SLmixing}Sigma-Lambda mixing on the lattice\protect\\}

Hadrons are studied on the lattice by calculation of correlation functions made from operators which are constructed to represent particular hadrons by incorporating their flavour content, symmetries and quantum numbers. The canonical way of doing this is to write down the operators for a particular hadron using the full SU(3)-flavour symmetry with definite isospin, since we know isospin to be a very good approximate symmetry in reality. 

In this study we wish to explore other SU(2) symmetries in addition to isospin and hence, to begin with, we do not appeal to the quark flavours up, down and strange, and instead use placeholders $a$, $b$ and $c$. 

\subsection{\label{sec:operators} Standard interpolating operators}

Following the notation introduced in \cite{Horsley:2014koa}, we employ standard Euclidean-space interpolating operators for the SU(3) $\Sigma^0$ and $\Lambda^0$ octet baryons with flavour content $a,b,c$,
\begin{eqnarray}
    \mathcal{B}_{\Sigma(abc),\alpha}(x) = \frac{1}{\sqrt{2}}\epsilon^{lmn}\big( b^l_\alpha(x)\left[ a^m(x)^{\top}C\gamma_5 c^n(x) \right]\nonumber\\ 
    +\,a_\alpha^l(x)\left[ b^m(x)^{\top}C\gamma_5 c^n(x) \right] \big),
\end{eqnarray}
and
\begin{eqnarray}
    \mathcal{B}_{\Lambda(abc),\alpha}(x) = \frac{1}{\sqrt{6}}\epsilon^{lmn}\big( 2c^l_\alpha(x)\left[ a^m(x)^{\top}C\gamma_5 b^n(x) \right]\nonumber\\ +\,b_\alpha^l(x)\left[ a^m(x)^{\top}C\gamma_5 c^n(x) \right] \nonumber\\ \,-a_\alpha^l(x)\left[ b^m(x)^{\top}C \gamma_5 c^n(x) \right] \big),
\end{eqnarray}
where $C=\gamma_2\gamma_4$, the superscript $\top$ denotes a transpose in Dirac space, $l$, $m$ and $n$ are colour indices and $\alpha$ a Dirac index. These interpolating operators are constructed to create states with definite iso-, U- or V-spin (see \cite{Meshkov:1963zz}) symmetry depending on the choice of Cartan subalgebra \cite{Georgi:1982jb} used in constructing the octet representation. The $\Sigma^0$ is symmetric in the flavours $a$ and $b$ (which define the SU(2) subalgebra) whilst the $\Lambda^0$ is anti-symmetric; this ensures that the $\Sigma^0$ and $\Lambda^0$ states are orthogonal when the $a$ and $b$ quarks are degenerate. From these operators we construct the matrix of correlation functions
\begin{eqnarray}
    C_{ij}(t) = \frac{1}{V_s}\textrm{Tr}_D\Gamma_{unpol}\left< \sum_{\vec{y}} \sum_{\vec{x}} \mathcal{B}_i(\vec{y},t) \bar{\mathcal{B}}_j(\vec{x},0)  \right>, \nonumber \\ \textrm{for} \quad 0 \ll t \ll T/2, \quad i,j=\Sigma(abc),\Lambda(abc), \label{eqn:corrmatrix}
\end{eqnarray}
where $\Gamma_{unpol} = (1+\gamma_4)/2$, $V_s$ is the spatial lattice volume and $T$ is the full temporal extent of the lattice. 
The correlation matrix is Hermitian and its diagonalization can hence be described by a single parameter which we call the mixing angle, $\theta_{\Sigma\Lambda}$. 

The labeling of the three distinct quark flavours $(abc)$ in the above agrees with the notation used in previous works \cite{Horsley:2014koa}, but for the purpose of this work it will sometimes be advantageous to promote these labels to an explicit functional dependence of the correlation matrix elements on the quark masses and charges: 
\begin{eqnarray}
    C_{ij}(t) \longrightarrow C_{ij}(t,m_a,m_b,m_c,e_a,e_b,e_c) \nonumber \\  \textrm{or} \quad C_{ij}(t,\vec{m}_{abc},\vec{e}_{abc}), \quad i,j=\Sigma,\Lambda,
\end{eqnarray}
where $\vec{m}_{abc}=(m_a,m_b,m_c)$ is used for brevity and similar for the quark electric charges. Note that the ordering of the labels, or explicit dependencies, is important, since the $\Sigma^0$ operator for example has a symmetry in the first two quark flavours, $a$ and $b$ in this case. In our notation this is indicated by the ordering of the labels. 

\subsection{\label{sec:extrapolation} Extrapolation scheme}

When we have degeneracy in the first two (distinct) quarks (i.e. degenerate masses and charges, and hence numerically identical propagators), by symmetry the correlation matrix is diagonal. Furthermore if the third quark is also degenerate (SU(3) symmetry), the correlation matrix is proportional to the identity. In general, the diagonal elements of the correlation matrix are symmetric under an interchange of the first two quarks, whilst the off-diagonal elements are anti-symmetric. 

For the case of a degeneracy between either the first and third or second and third quarks (these cases are simply related by the symmetry of the correlation matrix under interchange of the first two quarks) we find by explicit manipulations of the correlation functions that we are able to write
\begin{equation}
    C_{\Sigma(aba')\Sigma(aba')}  = \frac{\left( 3C_{\Lambda(aa'b)\Lambda(aa'b)} + C_{\Sigma(aa'b)\Sigma(aa'b)} \right)}{4} \label{eqn:su2rel1}
\end{equation}
\begin{equation}
    C_{\Lambda(aba')\Lambda(aba')}  = \frac{\left( 3C_{\Sigma(aa'b)\Sigma(aa'b)} + C_{\Lambda(aa'b)\Lambda(aa'b)} \right)}{4}
\end{equation}
\begin{eqnarray}
    C_{\Sigma(aba')\Lambda(aba')}  = \frac{\sqrt{3}}{4}\big( C_{\Lambda(aa'b)\Lambda(aa'b)} - \nonumber \\  C_{\Sigma(aa'b)\Sigma(aa'b)} \big). \label{eqn:su2rel3}
\end{eqnarray}
We have primed the second quark label $a$ to indicate that its flavour is distinct even though it is degenerate in terms of mass and charge, while the time dependence is left implicit. All correlation function relations in this subsection are assumed to be at equal times.

The relations given in equations~\ref{eqn:su2rel1}--\ref{eqn:su2rel3} can also be summarised as
\mathleft
\begin{equation}
    \begin{bmatrix} 
        C_{\Sigma(aba')\Sigma(aba')} & C_{\Sigma(aba')\Lambda(aba')} \\
        C_{\Lambda(aba')\Sigma(aba')} & C_{\Lambda(aba')\Lambda(aba')}
    \end{bmatrix}  = U [C_{i(aa'b)j(aa'b)}] U^T,
\end{equation}
\mathcenter
where the matrix $U$ is given by
\begin{equation}
    U = 
    \begin{bmatrix} 
        \frac{1}{2} & \frac{\sqrt{3}}{2}\\
        \frac{-\sqrt{3}}{2} & \frac{1}{2} 
    \end{bmatrix}. \label{eqn:Umatrix}
\end{equation}
These relations can also be reversed to write 
\begin{equation}
 C_{\Lambda(aa'b)\Lambda(aa'b)}  = \frac{\left( 3C_{\Sigma(aba')\Sigma(aba')} - C_{\Lambda(aba')\Lambda(aba')} \right)}{2}, 
\end{equation}
\begin{equation}
    C_{\Sigma(aa'b)\Sigma(aa'b)}  = \frac{\left( 3C_{\Lambda(aba')\Lambda(aba')} - C_{\Sigma(aba')\Sigma(aba')} \right)}{2} .
\end{equation}

If for example we associate the degenerate $a$ and $a'$ quarks with the up (down) quarks, and the $b$-quark with the strange, then equations~\ref{eqn:su2rel1}--\ref{eqn:su2rel3} give the U-spin (V-spin) correlators in terms of the isospin correlators. 

In order to construct an SU(3)-flavour breaking expansion in terms of quark masses and charges, we begin by supposing that we have three distinct quark flavours with no electric charge and degenerate mass $m_0$. 
We then give the quarks small charges, $\epsilon Q_i$, proportional to the $uds$ physical charges, $Q_i$, but scaled by the small parameter $\epsilon$ (so as to keep their ratios physical). 
Since the down and strange quarks are still degenerate (same charge), if we take the ordering of isospin ($uds$ here, however $dus$ is also isospin), the first order Taylor expansion in the quark charge parameter gives 
\begin{eqnarray}
C(\vec{m}_{uds,0},\epsilon \vec{Q}_{uds}) = C(\vec{m}_{uds,0},0) + \frac{S}{2}
    \begin{bmatrix} 
        1 & 0 \\
        0 & 1
    \end{bmatrix}\epsilon \nonumber\\+
    \frac{D_{\textrm{QED}}}{4}
    \begin{bmatrix} 
        - 1 & \sqrt{3} \\
        \sqrt{3} & 1
    \end{bmatrix}\epsilon,\hspace{1cm}
\end{eqnarray}
where $\vec{m}_{uds,0} = (m_0,m_0,m_0)$ and
\begin{equation}
S = \left(\frac{\partial C_{\Sigma\Sigma}(\vec{m}_{dsu,0},\epsilon \vec{Q}_{dsu})}{\partial \epsilon}+\frac{\partial C_{\Lambda\Lambda}(\vec{m}_{dsu,0},\epsilon \vec{Q}_{dsu})}{\partial \epsilon}\right)\Big|_{\epsilon = 0}
\end{equation}
\begin{eqnarray}
D_{\textrm{QED}} = \Bigg(\frac{\partial C_{\Sigma\Sigma}(\vec{m}_{dsu,0},\epsilon \vec{Q}_{dsu})}{\partial \epsilon}-\hspace{2cm}\nonumber\\
\frac{\partial C_{\Lambda\Lambda}(\vec{m}_{dsu,0},\epsilon \vec{Q}_{dsu})}{\partial \epsilon}\Bigg)\Big|_{\epsilon = 0}.
\end{eqnarray}
This form follows from realizing that with constant and equal mass parameters for the three quarks, the correlation matrix elements obey the relations in equations~\ref{eqn:su2rel1}--\ref{eqn:su2rel3} for all values of $\epsilon$, and hence the first derivatives of each isospin element can be written in terms of the diagonal U-spin elements. 

Since the term $C(m_{i,0},0)$ is proportional to the identity and has no effect on the eigenvectors, we make the definition
\begin{equation}
    C(\vec{m}_{uds,0},0) + \frac{S}{2}
    \begin{bmatrix} 
        1 & 0 \\
        0 & 1
    \end{bmatrix}\epsilon \equiv A\mathbb{I}_2,
\end{equation}
and also absorb a factor of $\epsilon$ into $D_\textrm{QED}$: $D_\textrm{QED}\rightarrow\epsilon D_\textrm{QED}$. We have now, to first order, described the breaking of SU(3) symmetry down to SU(2) U-spin symmetry in the $\Sigma^0$--$\Lambda^0$ correlation matrix by introducing electromagnetism. 
Next we seek to break the remaining SU(2) symmetry by expanding the correlation matrix in powers of $\delta m_i = m_i - m_0$, the deviations in the masses from the SU(3) symmetric point (QED aside) where all 3 flavours of quark have mass $m_i=m_0$. 
To do this we make use of equations~\ref{eqn:su2rel1}--\ref{eqn:su2rel3} to notice that, for example
\mathleft
\begin{equation}
    \frac{\partial C_{ij}(\vec{m}_{uds},0)}{\partial m_d}\Big|_{(\vec{m}_{uds},0)} = U_{ik} \frac{\partial C_{kl}(\vec{m}_{uds},0)}{\partial m_s}\Big|_{(\vec{m}_{uds,0})} U_{lj}^T,
\end{equation}
\mathcenter
for the matrix $U$ from equation~\ref{eqn:Umatrix}. We find that if we also enforce the constraint that the average quark mass is held fixed as we move away from the SU(3) symmetric point \cite{Bietenholz:2011qq}
\begin{equation}
    \bar{m} \equiv \frac{1}{3}(m_u+m_d+m_s) = m_0
\end{equation}
\begin{equation}
    \Rightarrow \delta m_u + \delta m_d + \delta m_s = 0
\end{equation}
\begin{equation}
    \textrm{for} \quad \delta m_i = m_i - \bar{m},
\end{equation}
then the first order expansion in the masses reduces to (see Appendix~\ref{appendix} for a more detailed calculation) 
\mathleft
\begin{equation*}
C(\vec{m}_{uds},\epsilon \vec{Q}_{uds}) = 
A\mathbb{I}_2 + 
    \frac{D_{\textrm{QED}}}{4}
    \begin{bmatrix} 
        - 1 & \sqrt{3} \\
        \sqrt{3} & 1
    \end{bmatrix} +
\end{equation*}
\mathcenter
\begin{equation}
    \frac{D_\textrm{QCD}}{4}
    \begin{bmatrix} 
        - 3(\delta m_u + \delta m_d) & \sqrt{3}(\delta m_u - \delta m_d) \\
        \sqrt{3}(\delta m_u - \delta m_d) & 3(\delta m_u + \delta m_d)
    \end{bmatrix}, \label{eqn:1storderqedqcd}
\end{equation}
where 
\mathleft
\begin{equation}
    D_{\textrm{QCD}} = \Bigg(\frac{\partial C_{\Sigma\Sigma}(\vec{m}_{uds},0)}{\partial m_s}-\frac{\partial C_{\Lambda\Lambda}(\vec{m}_{uds},0)}{\partial m_s}\Bigg)\Big|_{\vec{m}_{uds,0}}.
\end{equation}
\mathcenter
This first order expression is diagonalized to yield
\begin{equation}
    \tan{2\theta_{\Sigma\Lambda, \textrm{isospin}}} = -\sqrt{3}\left( \frac{D_{\textrm{QCD}}(\delta m_u - \delta m_d) + D_{\textrm{QED}}}{3D_{\textrm{QCD}}(\delta m_u + \delta m_d) + D_{\textrm{QED}}} \right). \label{eqn:Tmixing}
\end{equation}
Repeating this process with the starting point of U- and V-spin correlation matrices gives 
\begin{equation}
        \tan{2\theta_{\Sigma\Lambda, \textrm{V-spin}}} = \sqrt{3}\left( \frac{D_{\textrm{QCD}}(\delta m_u - \delta m_s) + D_{\textrm{QED}}}{3D_{\textrm{QCD}}(\delta m_u + \delta m_s) + D_{\textrm{QED}}} \right), \label{eqn:Vmixing}
\end{equation}
\begin{equation}
        \tan{2\theta_{\Sigma\Lambda, \textrm{U-spin}}} = \frac{\sqrt{3}D_{\textrm{QCD}}(\delta m_d - \delta m_s)}{3D_{\textrm{QCD}}(\delta m_d + \delta m_s) + 4D^\textrm{U-spin}_{\textrm{QED}}}, \label{eqn:Umixing}
\end{equation}
where $D^\textrm{U-spin}_{\textrm{QED}}$ parallels $D_{\textrm{QED}}$ but with  derivatives of correlation functions of definite isospin instead of U-spin.

Notice that if we set $D_{\textrm{QED}} = 0$, the parameter $D_\textrm{QCD}$ cancels and the isospin expression reduces to exactly that presented in leading-order $\chi$PT \cite{GASSER198277}, as well as the leading-order term presented in our previous work \cite{Horsley:2014koa} based on a more group-theoretic approach. It is interesting to note that the method used in \cite{Horsley:2014koa} was to diagonalize the octet mass matrix to find the mixing angle before expanding the $\Sigma^0$/$\Lambda^0$ masses in terms of quark masses, whilst we have herein expanded the correlation matrix in terms of quark masses and then diagonalized.

At this point it is necessary for us to recall that our parameters $D_i$ still carry an implicit time dependence, and unlike in the QCD-only limit, for full QCD+QED our mixing angle depends explicitly on these parameters. We can take the usual route of diagonalizing at large times in our lattice simulations where the ground state dominates the signal. As we will see in the results of Section~\ref{sec:results}, the time-dependence of the mixing angle appears to be much weaker than that of the effective masses of the baryons themselves.

\subsection{\label{sec:renorm} Running quark masses}

Since we are operating with QED, we must also consider that variation in the input bare quark masses will no longer result in the same mass differences for quarks of different charges, due to renormalization. 
This leads us to include one further parameter in our fit function, which comes from taking
\begin{equation}
    \delta m_u + \delta m_d + \delta m_s = 0 \rightarrow \frac{1}{Z_{2/3}}\delta m_u + \frac{1}{Z_{-1/3}}(\delta m_d + \delta m_s) = 0,
\end{equation}
where the $Z_i$ factors correct for the running of the masses due to QED, specifically to account for the resultant difference in the $u$ and $d/s$ quark propagators from an equal change in their respective quark mass parameters, analogous to the Dashen scheme presented in \cite{Horsley:2015vla}. 

Upon updating our fit functions and making the definitions 
\begin{equation}
        D_{\textrm{QED}} \rightarrow Z_{2/3}D_{\textrm{QED}} \quad \textrm{and} \quad Z \equiv \frac{Z_{2/3}}{Z_{-1/3}},
\end{equation}
we are left with 
\begin{equation}
    \tan{2\theta_{\Sigma\Lambda, \textrm{isospin}}} = \sqrt{3}\left( \frac{(\delta m_u - Z\delta m_d) + D_{\textrm{QED}}/D_{\textrm{QCD}}}{3(\delta m_u + Z\delta m_d) + D_{\textrm{QED}}/D_{\textrm{QCD}}} \right), \label{eqn:finalmixing}
\end{equation}
and similar for V-spin, whilst for U-spin no extra factor is needed explicitly, although in keeping with the formalism a factor $Z_{-1/3}$ can be thought to have been absorbed by $D^\textrm{U-spin}_{\textrm{QED}}$.

In practice, the mixing angle can be determined on the lattice by numerical diagonalization of the correlation matrix. By performing calculations at a range of quark mass parameters and constant electric charge parameters we are able to fit the parameters $D_{\textrm{QED}}/D_{\textrm{QCD}}$ and $Z$. Assuming the functional form of equation~\ref{eqn:finalmixing} we are able to use these parameters to extrapolate to the physical point - the details for which will be given in the following sections.

Finally, considering terms beyond leading order, we can incorporate some higher order terms by replacing

\begin{equation}
D_{\textrm{QED}} \rightarrow \Bigg[\sum_{n=0}^\infty \frac{1}{(n+1)!}\bigg(\epsilon\frac{\partial}{\partial \epsilon}\bigg)^n\Bigg]D_{\textrm{QED}},
\end{equation}
and absorbing all higher-order QED terms (however not mixed QED-QCD terms) into the existing parameter without changing our established functional form. This is possible because we do not wish to vary $\epsilon$ in our extrapolation to the physical point, and the correlation matrix exhibits U-spin symmetry for all values of $\epsilon$. We can therefore think of our parametrisation as including all orders of pure-QED terms automatically, in theory. In practice however, as will be discussed in Section~\ref{sec:results}, we scale our QED parameter linearly to match the physical electromagnetic coupling and hence negate the inclusion of these higher-order terms in the present study.

We have found that at second order in the quark masses two additional parameters must appear in the correlation matrix expansion, and further parameters for mixed QCD-QED terms. Given the relatively few number of lattice ensembles available in the current analysis we therefore forgo inclusion of higher order QCD terms.   

\section{\label{sec:lattice}Lattice scheme\protect\\}

\begin{table*}
\centering
\begin{tabular}{ |p{1.2cm}|p{1.7cm} p{1.7cm} p{1.4cm} | p{1.7cm} p{1.7cm} p{1.4cm}|p{1.9cm}|p{1.7cm}|}
 \hline
 \multicolumn{9}{|c|}{Lattice Ensembles} \\
 \hline
 volume & $\kappa_u, \, \kappa_d,\, \kappa_s \,$  & $(\textrm{sea})$ &  &  $\kappa_u,\, \kappa_d,\, \kappa_s \,$ & $(\textrm{valence})$ & & $\theta_{\Sigma\Lambda,\textrm{isospin}}$ & $M_{u\bar{u}}$ (MeV) \\
 \hline
 $24^3\times48$ & 0.124362 & 0.121713 & 0.121713 & 0.124362 & 0.121713 & 0.121713 & -30$^\circ$\ (theory) & 442(9) \\
 \hline
 $24^3\times48$ & 0.124374 & 0.121713 & 0.121701 & 0.124374 & 0.121713 & 0.121701 & -21.8(1.1)$^\circ$ & 423(9) \\
   &   &   &   & 0.124387 & 0.121713 & 0.121689 & -19.5(1.2)$^\circ$ & 423(10) \\
   &   &   &   & 0.124400 & 0.121740 & 0.121649 & -6(1)$^\circ$ & 378(28) \\
 \hline
 $24^3\times48$  & 0.124400 & 0.121713 & 0.121677 & 0.124400 & 0.121713 & 0.121677 & -17.8(7)$^\circ$ & 405(8) \\
   &   &   &   & 0.124420 & 0.121713 & 0.121657 & -16.7(7)$^\circ$ & 387(8) \\
   &   &   &   & 0.124430 & 0.121760 & 0.121601 & -4.8(7)$^\circ$ & 377(8) \\
 \hline
 $48^3\times96$ & 0.124508 & 0.121821 & 0.121466 & 0.124508 & 0.121821 & 0.121466 & -3.5(4)$^\circ$ & 284(4) \\
   &   &   &   & 0.124400 & 0.121713 & 0.121677 & -18.5(9)$^\circ$ & 389(5) \\
 \hline
\end{tabular}
\caption{Volumes, $\kappa$-values used in the generation of the lattice configurations, valence $\kappa$-values used in the calculation of the correlators on their respective backgrounds and the fitted isospin mixing angles. Physical electric charges were associated with the $\kappa$'s for each flavour of quark (although the coupling is non-physical; see Section~\ref{sec:results}) and we also present the lightest neutral flavour-singlet meson $M_{u\bar{u}}$ on each ensemble for reference. The mixing angle result for the first ensemble follows theoretically from equations~\ref{eqn:su2rel1}--\ref{eqn:su2rel3}. \label{tab:lattices}}
\end{table*}

We extract the $\Sigma^0$--$\Lambda^0$ mixing angles from a combination of $24^3\times48$ and $48^3\times96$, $N_f=1+1+1$, dynamical QCD+QED lattice simulations around the U-spin symmetric point (approximate SU(3)-symmetric point) defined in \cite{Horsley:2015vla}. The gauge actions used are the tree-level Symanzik improved SU(3) gauge action and the noncompact U(1) QED gauge action (further details in \cite{Horsley:2019wha,Horsley:2015vla}). The fermions are described by an $\mathcal{O}(a)$-improved stout link non-perturbative clover (SLiNC) action \cite{Cundy:2009yy}. The couplings used and lattice spacing are 
\begin{equation}
\beta_{\textrm{QCD}} = 5.5, \quad \beta_{\textrm{QED}}=0.8, \quad a^{-1}/\textrm{GeV} = 2.91(3),
\end{equation}
which gives a QED coupling $\alpha_\textrm{QED} \simeq 0.1$, roughly $10\times$ larger than the physical value.

We presently neglect electromagnetic modifications to the clover term.
  This will leave us with corrections of $O(\alpha_{QED} e^2 a)$, which turn out to be no larger than the $O(a^2)$ corrections from QCD (see \cite{Horsley:2015eaa} for numerical evidence of this).
  Adding an electromagnetic clover term with $c_{em} = 1$ would leave us with corrections of $O(\alpha_{QED} e^2 g^2 a)$ (to this order in $\alpha_{QED}$ ), which is not a significant improvement, if at all.
  Furthermore, since the current manuscript is concerned with isospin-breaking effects, these discretisation effects will be further suppressed by a power of $m_d - m_u$ for QCD effects and $e_u^2 - e_d^2$ for QED effects.

The lattice ensembles used for this study have been selected to focus on the region near an approximate SU(3)-flavour symmetry. Given the difference in charges, this symmetry cannot be exact, and our approach is to tune the neutral (connected) pseudoscalar mesons to be degenerate (see \cite{Horsley:2015vla}). Starting from this point, the approximate symmetry is further broken along a trajectory that leaves no residual invariant (or approximate invariant) SU(2) subgroup. In particular, we introduce a breaking $m_u-m_s$, while holding fixed both $m_d$ and the average quark mass. In this way we preserve the physical mass hierarchy,
$m_u<m_d<m_s$. To further improve the constraint on our expansion parameters we also consider partially-quenched (PQ) propagators, where the valence masses are allowed to vary independently of the simulated sea quarks. The simulation parameters used in this study are listed in Table~\ref{tab:lattices}.

\section{\label{sec:results}Results\protect\\}

From the expressions presented in equations~\ref{eqn:Tmixing}--\ref{eqn:Umixing}, in the absence of QED, it is clear that the mixing angle depends only on the relative quark mass splitting at first order,

\begin{equation}
        \tan{2\theta_{\Sigma\Lambda, \textrm{isospin},\textrm{QCD-only}}} = \frac{\delta m_d - \delta m_u}{\sqrt{3}(\delta m_u + \delta m_d)},
\end{equation}
provided that the average quark mass $\bar{m}$ is held constant. As a consequence, given a set of quark mass parameters, we can directly predict the QCD-only mixing angle. 

In a recent QCD-only lattice study of the $\Sigma^0$--$\Lambda^0$ mixing \cite{Horsley:2014koa}, the quark masses at the physical point were determined to be

\begin{equation}
    a\delta m_u = -0.01140(3), \quad a\delta m_d = -0.01067(3),
\end{equation}
giving
\begin{equation}
    \Rightarrow \theta_{\Sigma\Lambda, \textrm{isospin},\textrm{QCD-only}} = -0.55(3)^\circ.
\end{equation}

Note that we are using a sign convention that followed from our choice of ordering for isospin ($uds$ vs. $dus$) as well as the ordering of the $\Sigma^0$ and $\Lambda^0$ along the diagonal of the correlation matrix, and differs from that used in \cite{Horsley:2014koa}. 

We also note that using the mass ratios of the latest FLAG review \cite{Aoki:2019cca}, the QCD-only mixing angle is predicted to be $\theta_{\Sigma\Lambda, \textrm{isospin},\textrm{QCD-only}} = -0.65(3)^\circ$.

In the present work we directly determine the lattice mixing angle for each of the QCD+QED ensembles listed in Table~\ref{tab:lattices}. This is done by calculating all four elements of the $\Sigma^0$--$\Lambda^0$ correlation matrix (equation~\ref{eqn:corrmatrix}) on an ensemble, for each site in the time dimension of the lattice, and numerically diagonalizing the matrix $C(t)$. A constant is fitted to the observed plateau region. We perform this direct diagonalization in favour of the more typical generalized eigenvalue problem (GEVP, first introduced in \cite{Luscher:1990ck}) as it is more consistent with our extrapolation formalism, and the advantages of the GEVP are in extracting eigenvalues of $C(t)$, whilst we are herein only interested in the eigenvectors. 

Upon including QED, the resulting mixing angles are displayed in Table~\ref{tab:lattices}. These results are used to fit equation~\ref{eqn:finalmixing}.

\begin{figure}[b]
\includegraphics[width=0.48\textwidth,height=0.71\textwidth]{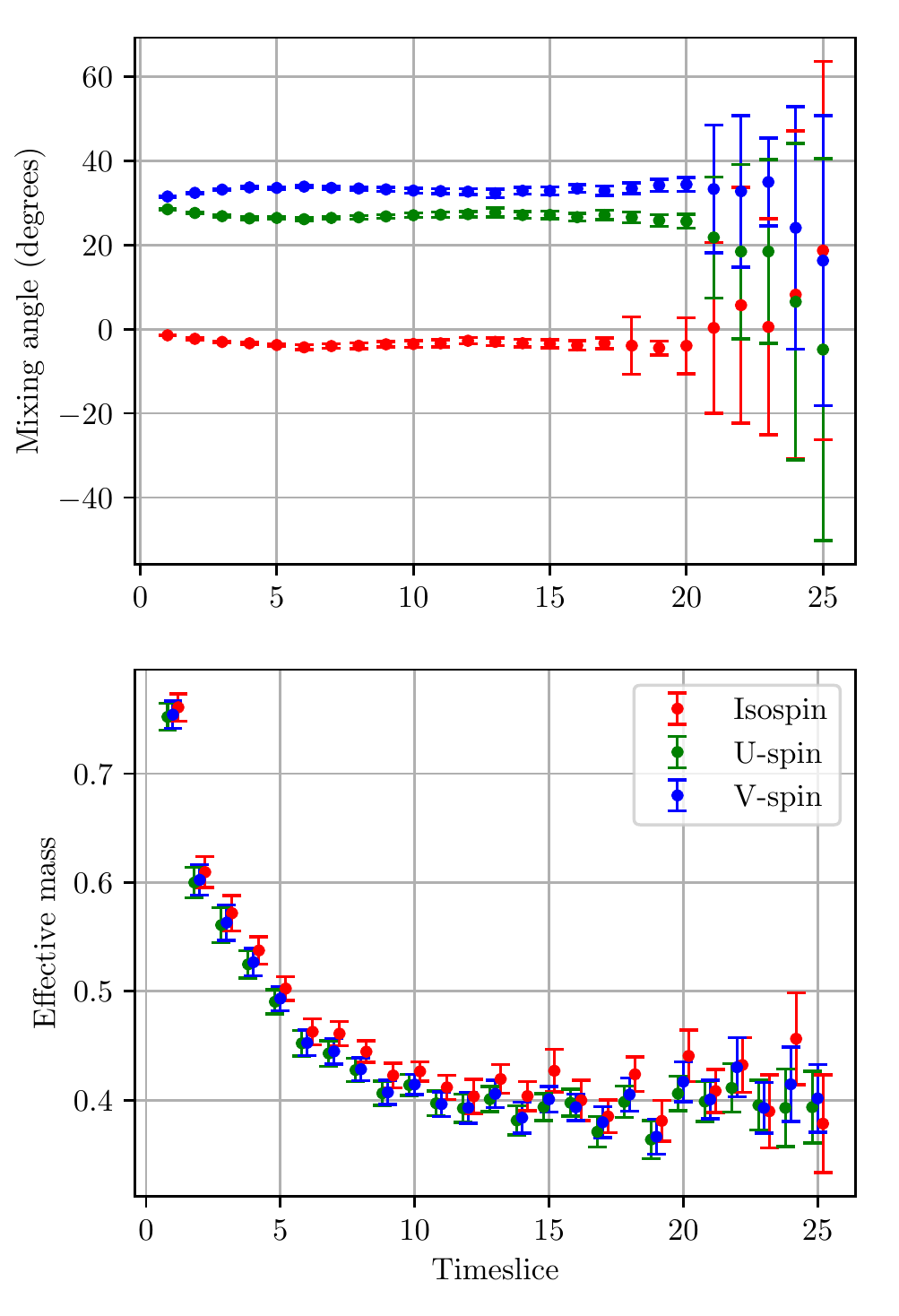}
\caption{An example of the mixing angle from diagonalization and corresponding effective mass, in this case on the largest-volume lattice used (unitary). The effective mass shows the $\Sigma^0$--$\Sigma^0$ component of the correlation matrix for each SU(2)-subgroup with a slight offset in time applied for clarity. \label{fig:anglefit}}
\end{figure}

It is a feature of our method for determining the mixing angle that we avoid fitting effective masses and instead fit the mixing angle directly. As it can be seen in Fig.~\ref{fig:anglefit}, the signal quality and ability to resolve the T-, U- and V-spin signals is much greater for the mixing angle, since it generally exhibits a much weaker time-dependence than the effective masses of the baryons themselves. Theoretically, we have shown that when QED is absent there is no time dependence in the mixing angle at first order, whilst this is no longer true for QCD+QED mixing, but as the QCD contribution to the mixing angle is much larger than the contribution of QED for most of our mass splittings, the mixing angle appears roughly time-independent. 

\begin{table}
\centering
\begin{tabular}{ |p{2cm}|p{3cm} p{1.6cm}|}
 \hline
 \multicolumn{3}{|c|}{Fit Parameters} \\
 \hline
    parameter & $D_{\textrm{QED}}$/$D_{\textrm{QCD}}$ & $Z$ \\
 \hline
    central value & $-3.8(7)\times10^{-5}$ & $0.96(4)$ \\
 \hline
\end{tabular}
\caption{The best-fit parameter values ($\chi^2/\textrm{DOF} = 0.84$) from the fit of the QCD+QED isospin mixing angles with $D_\textrm{QED}$ scaled to the physical EM coupling. The correlation coefficient for the two parameters is $-0.45$. \label{tab:params}}
\end{table}

\begin{figure}[b]
\includegraphics[width=0.48\textwidth,height=0.71\textwidth]{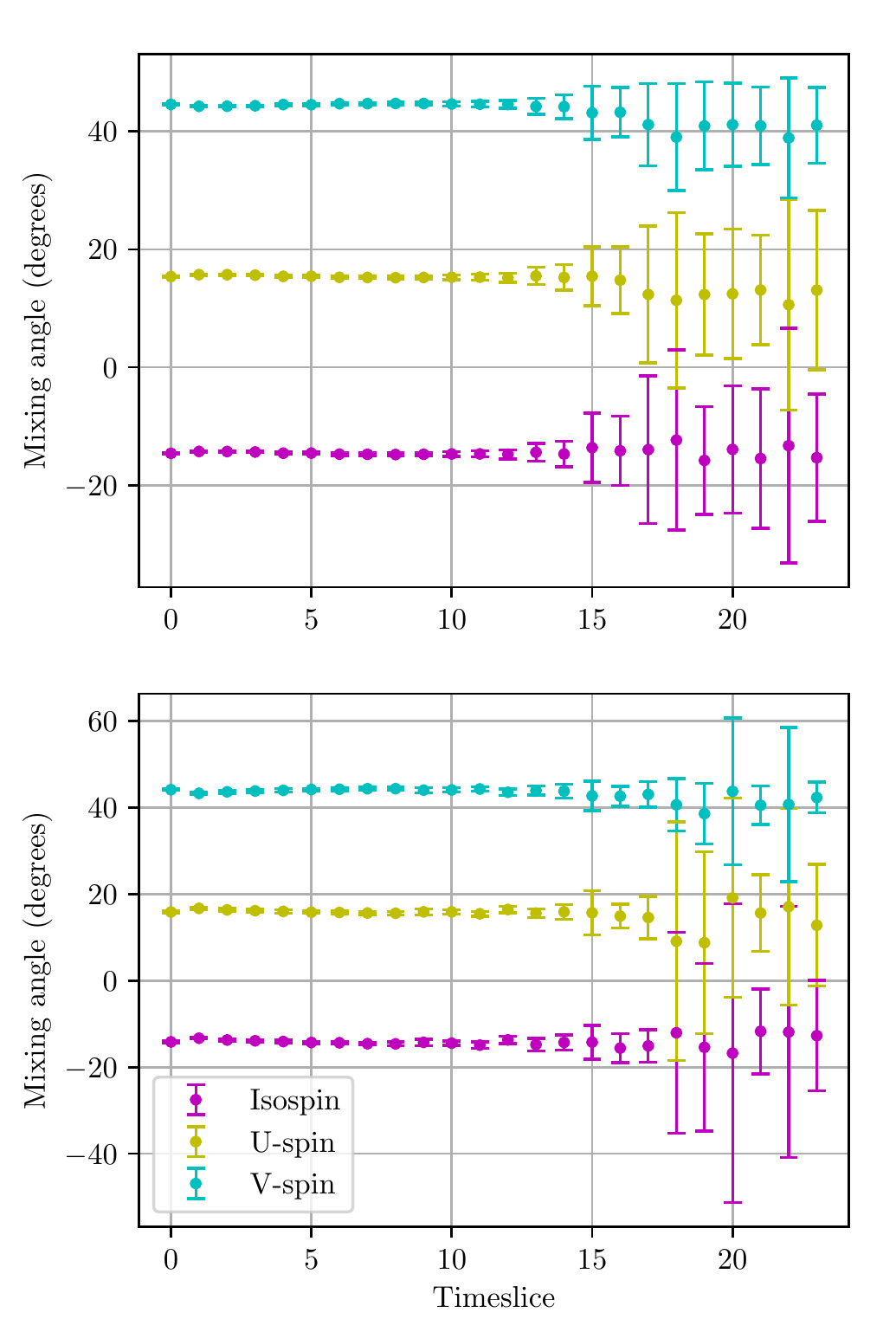}
\caption{This figure shows the T-, U- and V-spin mixing angles for the two PQ calculations we performed with the constituent quark charges set to zero. The top plot has $(\kappa_u,\kappa_d,\kappa_s)=(0.12092,0.1209,0.12088)$ and the bottom plot has $(\kappa_u,\kappa_d,\kappa_s)=(0.12094,0.1209,0.12086)$. The QCD-only mixing angle formula we have derived predicts that the mixing angles be time-independent at first order when QED is absent. \label{fig:neutangle}}   
\end{figure}

We have performed mixing angle calculations at two additional partially quenched points with the up, down and strange quark electric-charges set to zero which are presented in Fig.~\ref{fig:neutangle}, where we see our first-order prediction of time-independence of the QCD-only mixing angle (for constant $\bar{m}$) to be manifest.

\begin{figure*}
\includegraphics[width=.87\textwidth,height=0.68\textwidth]{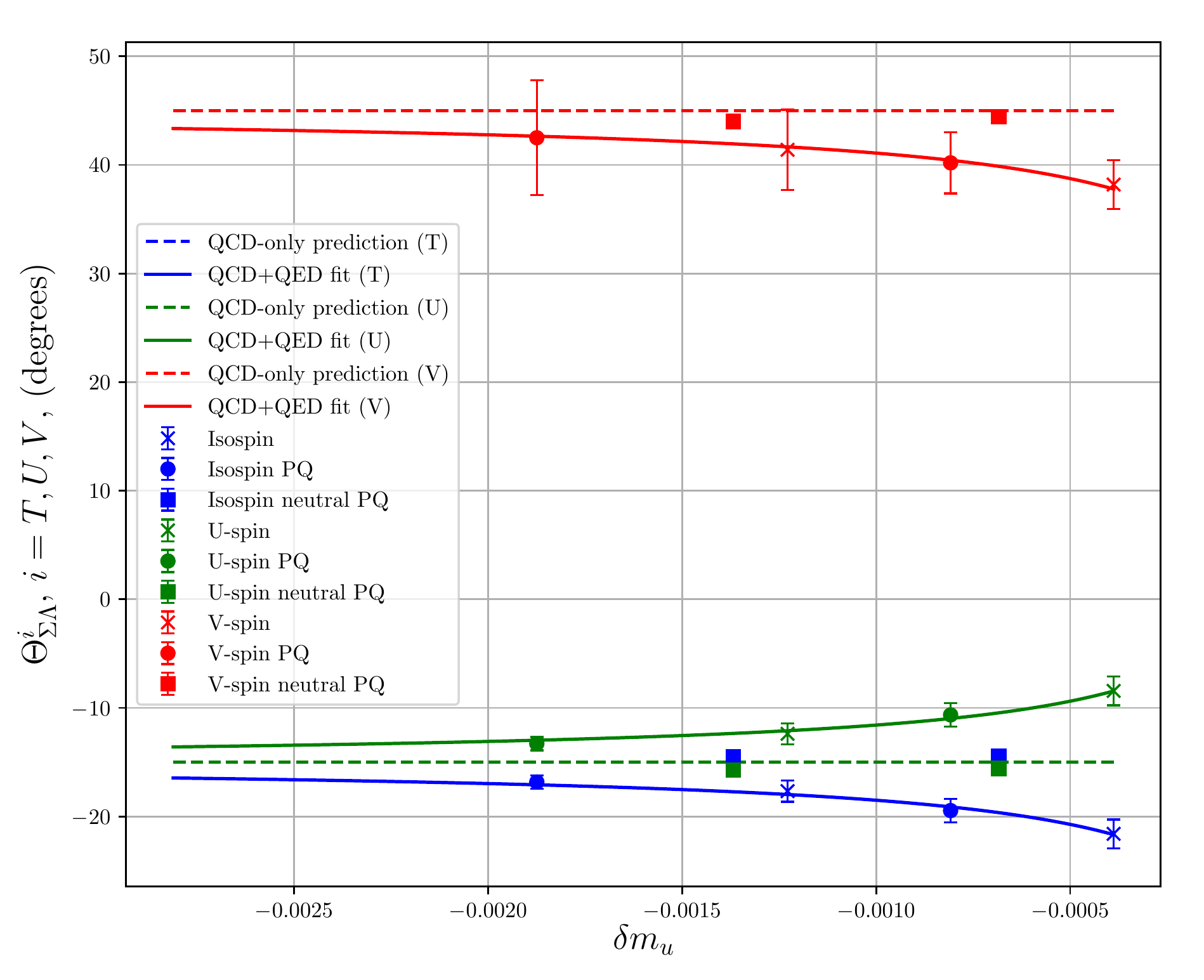}
\caption{This plot shows our simulation results for the mixing angles at the quark $\kappa$ values with $\delta m_d = 0$. In this scenario the QCD-only mixing angle is a constant (dotted lines) and we can see the QED-inclusive mixing angle asymptoting to that of only QCD as the T, U or V-spin symmetry becomes more broken by the mass parameters. The squares are mixing results from PQ calculations performed with all charges set to zero to approximate the QCD-only scenario whilst the circles are PQ with physical charges (unphysical coupling; see Section~\ref{sec:lattice}) and the crosses are unitary. \label{fig:qcdvqed}}
\end{figure*}

In Fig.~\ref{fig:qcdvqed} we show a comparison of the QCD-only and QCD+QED mixing angle fit-functions (equations~\ref{eqn:Tmixing}--\ref{eqn:Umixing} with $D_\textrm{QED}\equiv 0$ for QCD-only mixing) for lattice simulations with $\delta m_d = 0$, for which the QCD-only mixing angle function is constant. In addition to the relevant points from Table~\ref{tab:lattices}, we have included the two neutral partially quenched diagonalizations of Fig.~\ref{fig:neutangle}, which can be seen to agree well with the theoretical prediction of QCD-only mixing.

It can be seen in Fig.~\ref{fig:qcdvqed} and from our equations~\ref{eqn:Tmixing}--\ref{eqn:Umixing} that in the absence of QED ($D_\textrm{QED} \rightarrow 0$), the differences between any two of the mixing angles is a constant (magnitude $\pi/3$ or $\pi/6$), whilst with QED instated, only the isospin-V difference remains constant. 
This is because the isospin and V-spin doublets have the same combination of charges.
Another feature of Fig.~\ref{fig:qcdvqed} is that as we move to the left, and the SU(3)-flavour symmetry becomes more broken, the mixing angle for QCD+QED is asymptoting to that of QCD-only, which is a result of the bare quark mass differences becoming dominant in their mixing contribution over the electric charge differences (which do not change). 

As can be observed from Table~\ref{tab:lattices}, we have determined the mixing angle on both $24^3\times48$ and $48^3\times96$ volumes. The mixing angles on the larger volume are consistent with the $24^3\times48$ result and hence we do not attempt to correct for finite volume effects in this work. 
We have investigated the results of excluding the $48^3\times96$ result from our fits but given the small number of total simulations, it has proven to reduce the uncertainty by about $25\%$ to include it, despite the possibility of finite-volume differences.

To extrapolate our result to the physical point, we use the physical quark mass parameters determined in \cite{Horsley:2015vla}, 

\[
    a\delta m_u = -0.00834(8), \quad a\delta m_d = -0.00776(7),
\]
and

\[
    a\delta m_u = -0.00791(4), \quad a\delta m_d = -0.00740(4),
\]
using $32^3\times64$ and $48^3\times96$ volume lattices respectively. No $24^3\times48$ physical point is available, however our mixing angles seem consistent between $24^3\times48$ and $48^3\times96$ volumes within uncertainty, and it is interesting to observe the variation in the result due to systematic differences. Using the best fit parameters displayed in Table~\ref{tab:params}, where the parameter $D_\textrm{QED}$ has been scaled down by the proportionality factor $\alpha^\star_\textrm{QED}/\alpha_\textrm{QED}=0.07338$ that relates our simulated EM coupling $\alpha_\textrm{QED}$ to that of the real world, $\alpha^\star_\textrm{QED}$, we find

\[
    \theta_{\Sigma\Lambda, \textrm{isospin}}|_{\textrm{QCD+QED}} = -1.00(32)^\circ,
\]
and

\[
    \theta_{\Sigma\Lambda, \textrm{isospin}}|_{\textrm{QCD+QED}} = -0.96(31)^\circ,
\]
for the physical quark masses determined on $32^3\times64$ and $48^3\times96$ volumes respectively. These first-order results compare well with the widely used DvH formula result \cite{Dalitz:1964es,Gal:2015iha}, $-0.86(6)^\circ$, which also incorporates QED effects implicitly, in the sense that it cannot separate QCD from QED effects on the mixing angle. 

Whilst a direct confirmation of the validity of the assumed linear-in-$\alpha_\textrm{QED}$ scaling of the QED parameter has not yet been performed, it was shown in \cite{Cundy:2009yy} that $1/\kappa^c_q , 1/\bar{\kappa}_q$ and the bare quark mass at the symmetric point, $1/2\bar{\kappa}_q - 1/2\kappa^c_q$, all displayed linear behaviour with scaling of the quark charge squared.

We note that our renormalization parameter $Z$ is consistent with that presented in \cite{Horsley:2015vla} of $0.93$ using the Dashen scheme, which is defined by the running of connected neutral psuedoscalar meson masses. As was found in \cite{Horsley:2014koa}, the magnitude of the NLO QCD term was roughly one third that of the LO QCD term, and hence we approximate the contributions from higher order QCD terms as a systematic uncertainty of $20\%$. The effects of higher-order QED terms remains to be investigated.

\section{\label{sec:conclusion}Conclusion\protect\\}

In this work we have used the symmetry properties of the pure baryon-octet wavefunctions (and hence, interpolating operators) to simplify an expansion in QCD and QED parameters about the SU(3)-symmetric point, and consequently derived a scheme for extrapolating the $\Sigma^0$--$\Lambda^0$ mixing angle to the physical point along a path where the average quark mass is held constant.

We have observed that our extrapolation scheme accommodates past determinations of the mixing angle in the cases of both QCD-only mixing and physical QCD+QED mixing, and offers new insight into the interplay between QCD and QED effects on the mixing. We find that the QED contribution to the mixing angle at first order is of comparable magnitude to that of the quark mass differences and acts to effectively double the mixing.

\begin{acknowledgments}

The numerical configuration generation (using the BQCD lattice
QCD program \cite{Haar:2017ubh})) and data analysis
(using the Chroma software library \cite{Edwards:2004sx}) was carried out
on the IBM BlueGene/Q and HP Tesseract using DIRAC 2 resources
(EPCC, Edinburgh, UK), the IBM BlueGene/Q (NIC, J\"ulich, Germany)
and the Cray XC40 at HLRN (The North-German Supercomputer
Alliance), the NCI National Facility in Canberra, Australia
(supported by the Australian Commonwealth Government)
and Phoenix (University of Adelaide).
ZRK was supported by an Australian Government Research Training Program (RTP) Scholarship. 
RH was supported by STFC through grant ST/P000630/1.
HP was supported by DFG Grant No. PE 2792/2-1.
PELR was supported in part by the STFC under contract ST/G00062X/1.
GS was supported by DFG Grant No. SCHI 179/8-1.
RDY and JMZ were supported by the Australian Research Council Grants
FT120100821, FT100100005, DP140103067 and DP190100297.
We thank all funding agencies.
\end{acknowledgments}

\newpage
\appendix*
\section{\label{appendix}}

We present here a more detailed derivation of the QCD term in equation~\ref{eqn:1storderqedqcd} and note that the QED term is derived in an analogous manner. The starting point is the Taylor expansion in quark masses about the nominated SU(3)-symmetric point, $m_i=\bar{m}$. In the following we will ignore the correlation matrix dependence on the electric charges of the quarks, since they are identically set to zero throughout. We find, dropping $\mathcal{O}(\delta m^2)$ terms, 

\begin{widetext}

\begin{equation*}
    C_{ij}(\vec{m}_{uds}) \simeq  C_{ij}(\vec{m}_{uds,0}) +  \left(\frac{\partial C_{ij}(\vec{m}_{uds})}{\partial m_u}\Big|_{\vec{m}_{uds,0}}\right)\delta m_u  +  \left(\frac{\partial C_{ij}(\vec{m}_{uds})}{\partial m_d}\Big|_{\vec{m}_{uds,0}}\right)\delta m_d  +
    \left(\frac{\partial C_{ij}(\vec{m}_{uds})}{\partial m_s}\Big|_{\vec{m}_{uds,0}}\right)\delta m_s 
\end{equation*}

\begin{equation*}
     =  C_{ij}(\vec{m}_{uds,0}) + \left( U_{ik}^T \frac{\partial C_{kl}(\vec{m}_{uds})}{\partial m_s}\Big|_{\vec{m}_{uds,0}} U_{lj} \right) \delta m_u  + \left( U_{ik} \frac{\partial C_{kl}(\vec{m}_{uds})}{\partial m_s}\Big|_{\vec{m}_{uds,0}} U_{lj}^T \right) \delta m_d  + 
    \left(\frac{\partial C_{ij}(\vec{m}_{uds})}{\partial m_s}\Big|_{\vec{m}_{uds,0}}\right) \delta m_s,
\end{equation*}
or written as a matrix equation,

\begin{equation*}
    \begin{split}
    \Bigg[
    \begin{matrix}
         C_{\Sigma\Sigma}(\vec{m}_{uds,0}) + \frac{3}{4} \frac{ \partial C_{\Lambda\Lambda}(\vec{m}_{uds})}{\partial m_s}\Big|_{\vec{m}_{uds,0}}(\delta m_u + \delta m_d) + \frac{\partial C_{\Sigma\Sigma}(\vec{m}_{uds})}{\partial m_s} \Big|_{\vec{m}_{uds,0}}\left( \frac{1}{4}\delta m_u + \frac{1}{4}\delta m_d + \delta m_s \right) \\
        \frac{\sqrt{3}}{4}\left( \frac{ \partial C_{\Sigma\Sigma}(\vec{m}_{uds})}{\partial m_s} - \frac{\partial C_{\Lambda\Lambda}(\vec{m}_{uds})}{\partial m_s} \right)\Big|_{\vec{m}_{uds,0}}(\delta m_u - \delta m_d)
    \end{matrix} \cdots \hspace{2cm} \\ \cdots
    \begin{matrix}
        \frac{\sqrt{3}}{4}\left( \frac{ \partial C_{\Sigma\Sigma}(\vec{m}_{uds})}{\partial m_s} - \frac{\partial C_{\Lambda\Lambda}(\vec{m}_{uds})}{\partial m_s} \right)\Big|_{\vec{m}_{uds,0}}(\delta m_u - \delta m_d)  \\
         C_{\Sigma\Sigma}(\vec{m}_{uds,0}) + \frac{3}{4}\frac{ \partial C_{\Sigma\Sigma}(\vec{m}_{uds})}{\partial m_s}\Big|_{\vec{m}_{uds,0}}(\delta m_u + \delta m_d) + \frac{\partial C_{\Lambda\Lambda}(\vec{m}_{uds})}{\partial m_s} \Big|_{\vec{m}_{uds,0}}\left(\frac{1}{4}\delta m_u + \frac{1}{4}\delta m_d + \delta m_s \right)
    \end{matrix}\Bigg],
\end{split}
\end{equation*}
and this, upon making the constant-$\bar{m}$ substitution $\delta m_s = -\delta m_u - \delta m_d$, reduces to 

\begin{equation*}
    C(\vec{m}_{uds}) \simeq 
    C(\vec{m}_{uds,0})
    +
    \frac{1}{4}
    \begin{bmatrix} 
        - 3(\delta m_u + \delta m_d) & \sqrt{3}(\delta m_u - \delta m_d) \\
        \sqrt{3}(\delta m_u - \delta m_d) & 3(\delta m_u + \delta m_d)
    \end{bmatrix}
    \left( \frac{ \partial C_{\Sigma\Sigma}(\vec{m}_{uds})}{\partial m_s} - \frac{\partial C_{\Lambda\Lambda}(\vec{m}_{uds})}{\partial m_s} \right)\Big|_{\vec{m}_{uds,0}},
\end{equation*}

\end{widetext}
and we can now see the connection to equation~\ref{eqn:1storderqedqcd}, with the term in parentheses being the previously defined $D_\textrm{QCD}$. The above expression can be directly diagonalized to yield the QCD-only mixing formula familiar from $\chi PT$ (for constant-$\bar{m}$) and the matching first order term presented in \cite{Horsley:2014koa}. Furthermore we can make the connection to our previous work that for the expansion parameter $A_2$ in \cite{Horsley:2014koa}, at large times we must have $ D_\textrm{QCD}(t) = 2 A_2 $.

\bibliography{qcdpapers}

\end{document}